\documentstyle[preprint,aps]{revtex}
\begin{document}

\draft

\title{Pre-equilibrium evolution of non-Abelian plasma}

\author{Gouranga C Nayak \cite{e1} and V. Ravishankar \cite{e2}}
\address{Department of Physics, Indian Institute of technology, 
Kanpur -- 208 016, INDIA }

\maketitle

\begin{abstract}

 We study the production and the equilibration of a
 non-Abelian 
 $q\bar{q}$ plasma in an external chromoelectric field,
 by solving the Boltzmann equation with the non-Abelian
 features explicitly incorporated.
 We consider the gauge group $SU(2)$ and show that the
 colour degree of freedom has a major and dominant role
 in the dynamics of the system. It is seen that the
 assumption of the so called Abelian dominance is not
 justified. Finally, it is also shown that many of the
 features of microscopic studies of the system appear
 naturally in our studies as well.

\end{abstract}


\section{INTRODUCTION}
 There is a high expectation that with the advent of new 
 machines (RHIC at BNL, LHC at CERN) it will be possible to 
 study
 the quark gluon phase of strongly interacting matter. Lattice
 QCD predicts that at density and temperature of the order of
 $2GeV/fm^3$ and 200 MeV respectively, hadronic matter 
 undergoes
 a transition to the quark-gluon phase \cite{ann rev}; 
 the Au-Au collisions at 
 200 GeV/nucleon (in the centre of mass frame) in RHIC 
 are expected 
 to provide such energy densities which would then 
 give rise to the quark-
 gluon phase \cite{shukraft}. It is also generally 
 believed that the transition is of 
 the first order, although more recent lattice studies 
 \cite{jacob} and those of 
 intermittency \cite{hwa}
 suggest that this belief needs more substantiation.

 The most important question experimentally is
of course the signature of quark gluon plasma (QGP). 
Many proposals have been made, prominent
among them being $J/{\Psi}$ suppression \cite{matsui}, 
strangeness enhancement \cite{rafelski} and
dilepton production \cite{strickland}. Clearly, it is 
important to know, even if
only roughly, the dynamics of the QGP evolution in 
space - time. Indeed, apart from being intrinsically
interesting, the space - time   
dynamics will have a direct impact on the signatures proposed.

The space time evolution of QGP that is produced in
ultra-relativistic heavy ion (URHIC) collisions can be 
described \cite{Jacob} broadly in four stages.
They are, i) pre-equilibrium, ii)equilibrium, iii) cooling,
and iv) hadronisation. While most of the earlier work refers
to
stage (ii) where one studies the hydrodynamic 
evolution \cite{baym}, subsequently
there has been a good deal of effort in understanding 
stage (i) as well.
Since it is of relevance to us here, we briefly reveiw 
them below.

In the context of Flux-tube model (described in the 
next section), the 
transport equation is solved to study the evolution of 
the system. Assuming
the Bjorken scenario \cite{bjorken} where a central
rapidity
plateau will be seen in relativistic 
heavy ion collisions \cite{LHC},
this model deals with a
baryon free plasma in the central region; there would 
initially
be a huge deposit of energy which then creates the partons. 
In the approach
of Baym \cite{baym}, these partons are assumed to be 
created 
soon after the two nuclei have collided with each other,
and are further assumed to equlibrate almost instantaneously
 thereafter (with a very small relaxation time). This
 quark-gluon gas then proceeds to expand hydrodynamically.
Subsequent work of Kajantie and Matsui \cite{kajantie},
incorporated a dynamic particle production 
by introducing a source term  via the Schwinger mechanism
\cite{schwinger} 
in the Boltzmann equation, and they  
also studied the equilbration in the relaxation time 
approximation. 
On the other hand, Bialas {\it et al} \cite{bialas}
generalized the Baym analysis by introducing the effect of 
the 
background field on
the otherwise hydrodynamic flow. A self consistent 
study of the system was carried out by
Banerjee {\it et al} \cite{banerjee} who combined the 
effect of both the background field and the collisions
on the quark-antiquark system. 
 Note that in the analyses of Kajantie {\it et al} 
 \cite{kajantie} and 
 Banerjee {\it et al} \cite{banerjee}, the source term 
 acquires a time dependence by virtue of the time 
 dependence that
 the electric field suffers because of the particle
 production.
 Finally, by employing the same analysis as 
 of Banerjee {\it et al}, 
 Asakawa and Matsui \cite{asakawa} have studied the 
 rate of dilepton production by $q\bar{q}$
 annnihilation, as a function of (proper) time.
 All the above analyses ignore the gluon component.

In contrast to the above studies which we shall
broadly term macroscopic, there is an interesting 
and a more complete as well as a
complementery approach by Geiger and Kapusta \cite{geiger}.
They have 
studied  the QGP evolution in what they call the parton 
cascade
model. In this approach, the production and the evolution
of q$\bar{q}$ pairs and gluons are studied numerically, by
considering the direct collision processes 
$2 \rightarrow 2$ , the inelastic fusion processes 
$2 \rightarrow 1$, and the decay processes 
$1 \rightarrow 2$.
 The above processes are studied within the frame work
of the perturbative QCD, by taking the leading order 
and the next to
leading order contributions.
While this approach of Geiger and Kapusta has the merit 
of being
microscopic, it comes with the complication of extensive 
numerical
computation - and the physical picture may not always be 
clear.
More significantly for our purposes, the analysis 
confirms the Bjorken
hypothesis of the boost invariance of the parton densities 
in the
central rapidity region. Further, the relaxation time 
which is other-wise
free in the flux-tube models also gets determined. 
Indeed, Geiger et al \cite{geiger} report a value of 
$\sim  0.2 fm$ in their paper.
One may thus
look upon the flux tube models to be effective versions 
of the more detailed
microscopic pcm, if the $\tau_c$ determined by the 
latter is employed in
the transport equation. We shall make a more detailed 
comparison in the
subsequent sections.

In any case it is our purpose to emphasize and study
a crucial feature of the quark-gluon phase, {\it viz}, 
the inherently non-
Abelian nature of the plasma. Note that all the earlier 
macroscopic 
approaches \cite{baym,kajantie,bialas,banerjee,asakawa}
are in the so called Abelian approximation. Put simply,
the non-Abelian features are completely ignored in the 
space time evolution, 
and the Boltzmann equation describes the dynamics which 
is essentially that of a Maxwell fluid, be it 
in the source term or
in the background field term. The approach of Geiger and 
Kapusta \cite{geiger}
does incorporate the non-Abelian features in their basic 
Feynman
diagrams, but is neverthless incomplete in that 
the distribution function in its final form
does not carry any color degee of freedom.

We attempt to fill this gap, within the
framework of transport equations, by explicitly 
incorporating the
non-Abelian features mentioned above. To be sure, the 
relevant formalism
has already been devloped by Heinz \cite{heinz} . 
Here one needs to carefully
implement the non-Abelian constraints within the Boltzmann 
equation
such that the Bjorken picture of boost invariance 
\cite{bjorken} 
along the collision
direction is also preserved.

Before we proceed to present the formalism it
is pertinent to make the following observations: 
1) Strictly speaking,
the source term that has been employed in the above studies 
is inadmissiible.
 For, even within the so called Abelian approximation, 
 the chromoelectric 
field suffers a space time evolution in any 
self-consistent calculation.
To wit, the production of particles is then strictly 
governed by a time
dependent field, in which case a perturbative 
mechanism will take over
from the non-perturbative Schwinger mechanism. Note that the 
latter holds only for a uniform
constant field. All the previous studies employing the 
flux tube model
employ the non-perturbative expression. 2) As already
mentioned, the source term is derived for QED processes 
and needs to
be redone for QCD processes at hand. It is apparent 
that the QCD effect 
will show up most manifestly in the gluon production; there is
no corresponding counterpart in QED. In other words it is 
necessary to obtain a source 
term for gluons. 3) It is well known from earlier studies 
of Yang-Mills  equations
that there are inequivalent gauge field configurations 
which yield the
same field tensor \cite{wu}, and hence the same 
energy momentum tensor. 
This feature which is again absent in the Maxwell case 
has to be 
taken into account
in our study. Recall that the lattice studies at any 
given energy 
density implicitly sum over the contributions coming 
from all such 
inequivalent configurations \cite{fnotegavai}. 4) Finally,
 the inherent non-linearity and the local gauge
 invariance which characterize the Yang-Mills equations
 lead to an added complexity; the equations do not
 possess a unique solution, even after gauge fixing, and
 even if one considers only static configurations.
 Consequently, one needs to scan, with some suitable
 weightage, all the solutions for the given initial
 conditions, and sum over all such configurations. It
 is not clear if any procedure is known that allows an
 implementation of this requirement.

In this first study we restrict ourselves to 
a study of the simplest of the situations leaving the 
points mentioned 
above for a later presentation. Apart from acting as 
a warm up, it will
also allow us to devise a simple bench mark that would 
enable us to compare
the Abelian and the non-Abelian features. With that in mind 
we consider
only Maxwell like fields (see below), ignore the production 
due to a
time dependent field and, finally restrict ourselves to the 
gauge 
group $SU(2)$. The
last choice leads to an additional simplification as 
quarks, antiquarks
and gluons all belong to the same represantation. As a 
final simplification,
we shall also drop the gluonic terms altogether. Please 
note that 
even with all these approximations (which we shall remedy in 
later publication) our
approach is none-the worse compared to the fully 
Abelian models which
have been employed so far.

The plan of the paper is as follows.
In the next section, we shall briefly describe the model where we 
set up the transport equation
with the non-Abelian features appropriate to the 
group $SU(2)$ explictly incorporated. We
also obtain the auxiliary equation that follows from the 
requirement of the conservation of
the energy momentum tensor. Section {\bf 3} describes briefly
the numerical procedure employed. In Section 4 we present
the results and discussions. We conclude and summarize
the main results in Section 5.

\section{The Flux Tube Model}

One well known signature for the QGP is the dilepton 
production \cite{strickland}
which has the attractive feature that the leptons do 
not suffer any final state
(strong) interaction. As such this process can 
therefore be used to study
the space-time evolution of the QGP. In particular, one 
would be able to study
the pre-equilibrium regime, provided one can determine 
the distribution 
functions
for the quarks and the anti-quarks at such early times.

The flux tube model, as we have mentioned, is a useful 
choice in this context. We shall employ this model in
this paper. A brief description of the Flux Tube model
is not out of place here.

 In this model the two nuclei that undergo a 
 central collision at
ultra high energies are lorentz contracted  as thin plates 
\cite{low,nussinov}. These two
lorentz contracted nuclei pass through
each other and, in the process, acquire a non-zero colour 
charge ($\langle Q \rangle =
0, \,\langle Q^2 \rangle \neq 0$), by exchanging soft gluons. 
So one may figuratively call such nuclei
after collision as color capacitor plates between which a 
strong chromoelectric field is created. 
This field creates q$\bar{q}$ and gluon pairs {\it via}
the Schwinger mechanism 
which enforces the instability of the vacuum in the presence 
of an external electric field.
The partons so produced will collide with each other and also 
get accelerated by the
parent background field. The mutual collisions drive the 
system 
towards equilibrium
with suitable modulation from the background acceleration.

Let $f(\bar{f})$ be the one particle(anti-particle) 
distribution 
in the phase space.
In the earlier Abelian approximation, the evolution of 
the particles 
is studied via the 
Boltzmann-Vlasow equation, 
	
\begin{equation}
\left[ p_{\mu} \partial^\mu \pm g F_{\mu\nu} p^\nu 
\partial^\mu_p
\right]  f(x,p) (\bar{f}(x, p))=C(x,p)+S(x,p),
\label{two}
\end{equation}

where $g$ is the strength of the charge which couples to
the field.  Note that the distribution functions 
$f$, $\bar{f}$ 
are defined in a six
dimensional phase space of coordinate and momenta. In the
above equations, $C$ is
the collision term driving the system into equilibrium, 
and $S$ is the term that acts as the
source for pair production \cite{schwinger}.
We now generalize Eqn. (1) to incorporate the non-Abelian 
features explicitly.
An easy way to to accomplish this 
is to extend the phase-space by taking the color degrees of 
freedom into account. 
The extended phase space is now a direct sum
${\cal R }^6 \oplus {\cal G}$,
where ${\cal G}$ is the (compact) space 
corresponding to the given
gauge group.
Thus the single particle phase space has a dimension 
$d=6+ (N^2-1)$  if we consider  $SU(N)$. 
The evolution of the color charge now follows Wong's 
equation \cite{wong}

\begin{equation}
{dQ^a\over d{\tau}}= f^{abc} u_\mu Q^b A^{c\mu} 
\label{one}
\end{equation}

 which supplements the Lorenz force equation

 \begin{equation}
 {dp^{\mu} \over d{\tau}} = Q^a F^{a\mu\nu}u_{\nu}
 \end{equation}

 where $u_{\mu}$ is the four-velocity and $f^{abc}$ are the 
structure constants of the
  gauge group.
For SU(2) charges that we are interested in, the non-Abelian 
extension of Eqn.(1) reads 

\begin{equation}
\left[ p_{\mu} \partial^\mu + Q^a F_{\mu\nu}^a p^\nu 
\partial^\mu_p
+ \epsilon^{abc} Q^a A^b_\mu
p^{\mu} \partial_Q^c \right]  f(x,p,Q)=C(x,p,Q)+S(x,p,Q)
\label{two}
\end{equation}
 
 Note that we do not have to write a separate equation 
 for anti-quarks since 
 they belong to the same color representation as quarks. 
 Indeed, the antipodal points on the sphere (corresponding to 
the color part
 of the phase-space with a fixed charge) represent the 
 particle and 
 anti-particle.
 Eqn. (3) may be recast in the more convenient form 

\begin{equation}
{d \vec{Q} \over d \tau} = {u_\mu \vec{Q} \times \vec{A}^\mu}
\end{equation}
       
where the arrows now denote the direction in the color 
space. In the same 
notation,the transport equation obtains the form

\begin{equation}
\left[ p^\mu\partial_\mu+\vec{Q}\cdot \vec{F}_{\mu\nu}
p^{\mu}\partial_p^\nu
+p^{\mu}\vec{Q}\times \vec{A}_\mu \cdot 
{\partial \over \partial \vec{Q}} 
\right]
f(x,p,\vec{Q})
=C(x,p,\vec{Q})+S(x,p,\vec{Q})
\label{fourteen}
\end{equation}

For the model at hand, let us take the `plates' to be 
moving along 
the z-direction. 
As mentioned earlier, we restrict ${\vec{F}^{\mu\nu}}$ to 
be Maxwell like, 
{\it i.e.}, we restrict to only that field configuration
for which there exists a gauge choice such that the gauge
potentials commute with each other every where.
We next require a boost invariant description of the 
distribution functions as well as the other physical
quantities which may be determined thereof, even as the
system is evolving.
Keeping this in mind, 
we make the gauge choice where only the components
$A^{\mu a}=(A^{03},A^{33})$
are non-vanishing and the other components are zero.
 With this choice, the resulting electric field points 
 in the `3' direction
in the colour space. 
We now proceed to impose the Lorentz gauge condition, 
which has to be done such 
that the chromoelectric field depends
on the boost-invariant(along the axis of collision) quantity 
$\tau =(t^2-z^2)^{1/2}$. 
Observing that the system is effectively (1+1) dimensional, 
in the t-z plane, we write 

\begin{equation}
\vec{A}^{\mu}=\epsilon^{\mu\nu} \partial_{\nu} \vec{G}(\tau)
\label{thirteen}
\end{equation}

where  the indices $\mu ,\nu$ take values 0, 3 and $\vec{G}$ 
is a Lorentz scalar 
function depending only on $\tau$.
The above choice automatically implements the Lorentz gauge 
condition $\partial_\mu \vec{A}^\mu=0$.
 Clearly, the chromoelectric field $\vec{E}$ is 
dependent only on 
$\tau$, and is given by

\begin{equation}
\vec{E}(\tau)=\left[ {d^2 \over d\tau^2} -
(2/\tau){d \over d\tau} \right] \vec{G}(\tau)
\end{equation}

The Wong equation guarantees the conservation of the 
magnitude 
of the vector charge $\vec{Q}$, 
which may now be held fixed. Being the analogue of the 
Larmor equation for a charged
particle in an external magnetic field, it also conserves the 
component of the charge
that is parallel to the external chromoelectric field. 
It is therefore convenient to resolve the SU(2) charge in the 
polar coordinates.
Writing	

\begin{equation}
Q_1=Q sin\theta cos\phi; \, \, Q_2=Q sin\theta sin\phi ;\,
\,
Q_3=Q cos\theta \, ,
\label{thirteen}
\end{equation}
       
it is straight forward to verify that
      
\begin{equation}
\left( \vec{Q} \times { \partial  \over \partial \vec{Q} }
\right)_3 
={\partial  \over \partial \phi}
\end{equation}
     
The last equation will lead to considerable simpification 
in solving the transport
equation.
       
Let us now consider the collision term. As mentioned 
in the introduction, the colour
plate model can be looked upon as an effective version 
of the more detailed
parton cascade model which has confirmed the 
Bjorken hypothesis. We may therefore,
employ the simple relaxation time hypothesis, with the 
phenomenological
parameter $\tau_c$, the relaxation time, to be 
obtained from microscopic computations.
It then follows that

\begin{equation}
C = {-{p^\mu u_\mu (f-f_{eq})} \over {\tau_c}}
\end{equation}
       
where $f_{eq}$ is the eqilibrium distribution function,
 with local (space time dependent)values of the
 thermodynamic quantities. While it has been customary to
 take $f_{eq}$ to be one of an ideal gas, with some
 support from the pcm analysis of Geiger and Kapusta
 \cite{geiger}, recent lattice studies  
 suggest
 that the quark-gluon phase is possibly a
 non-perturbative phase. Since we are not dealing with
 a true system here, we shall conveniently take $f_{eq}$
 to be that of an ideal Fermi gas with a local temperature,
 evolving as a function of the proper time $\tau$. We
 thus have

\begin{equation}
f_{eq} = \frac{2}{\exp (p^\mu u_{\mu} /T(\tau))+1},
\label{six}
\end{equation}
   
where $u^\mu$ is the flow velocity,

\begin {equation}
u^\mu=(cosh\eta,0,0,sinh\eta).
\label{four}
\end{equation}
	
which is written in terms of the (space-time) rapidity
$ tanh\eta=z/t$.

Now we demand the boost invariance following 
Bjorken's picture according to which the longitudinal 
boosts are the 
symmetry operations on the single particle distribution. 
The boost invariant parameters
on which $f$ can depend are, apart from the charge
coordinates,

\begin{equation}
\tau =(t^2-z^2)^{1/2},\, \xi=(\eta-y) ,\, 
p_t=(p_0^2-p_l^2)^{1/2}
\end{equation}

where $y=\tanh^{-1}(p_l/p_0)$ is the momentum rapidity.
It is convenient to write separate equations for quarks
and anti-quarks, {\it a la} the Abelian case. If we
therefore identify the quark states with the points on the
upper hemisphere of the colour sphere, and the antiquarks
with the lower hemisphere, by a trivial relabeling, we may
write two equations,

 \begin{eqnarray}
\left[ {\partial  \over \partial \tau}
-\left( {\tanh \xi \over \tau}  + {g \cos \theta 
E(\tau) \over p_t \cosh \xi} \right) 
{\partial  \over \partial\xi}
+ g {d \over d\tau}G(
\tau)\tanh\xi{\partial \over \partial\phi}\right] 
f(\tau,\xi,p_t,\theta
,{\phi}) \\
+ {f \over \tau_c} = {f_{eq} \over \tau_c} + {\Sigma (\tau,
p_t,\xi,\theta) \over p_t \cosh \xi}
\label{eighteen}
\end{eqnarray}

\begin{eqnarray}
\left[ {\partial  \over \partial \tau}
-\left( {\tanh \xi \over \tau} -  {g \cos \theta 
E(\tau) \over p_t \cosh \xi} \right) 
{\partial  \over \partial\xi}
+ g {d \over d\tau}G(
\tau)\tanh\xi{\partial \over \partial\phi}\right] 
{\bar f}(\tau,\xi,p_t,\theta
,{\phi}) \\
+ {{\bar f} \over \tau_c} = {f_{eq} \over \tau_c} + 
{\Sigma (\tau,
p_t,\xi,\theta) \over p_t \cosh \xi}
\label{eighteen}
\end{eqnarray}

with the first of them for the quarks and the next 
for the anti-quarks. The angle variable $\theta$ varies
from 0 to $\pi/2$ in both the equations.
Finally, $\Sigma$ is the (non perturbative) 
Schwinger's expression for 
pair production and is given by

\begin{equation}
\Sigma(\tau,\xi,p_t,\theta)=-\frac{gE \cos\theta}{8\pi^3} 
\ln \left[ 1-\exp \left(
-{2\pi p_t^2 \over gE \cos\theta} \right) \right] 
({\alpha \over \pi})^{1/2}
\exp (-\alpha\xi^2)
\label{ninenteen}
\end{equation}
	
where we have inserted the Gaussian dependence on 
$\xi$ by hand. Note that the non-Abelian nature features
dominantly {\it via} the $cos(\theta)$ term. This
occurence plays an important role in the evolution of the
system.
	
After having set up the relevant equations 
(not all yet since energy-momentum
conservation is to be imposed), we observe that
the above differential equation possesses 
the formal solution

\begin{equation}
f(\tau,\xi,p_t\theta,\phi)=\int^\tau_0 d \tau^\prime 
\exp (\frac{\tau^\prime-
\tau}{\tau_c}) \left[ \frac{\Sigma 
(\tau^\prime,\xi^\prime,p_t,\theta )}
{p_t \cosh \xi^\prime} + \frac{f_{eq}
(\tau^\prime,\xi^\prime,p_t)}
{\tau_c} \right]
\label{twelve}
\end{equation}

where $\xi(\tau\prime)$ is given by

\begin{equation}
\xi^\prime= \sinh^{-1} \left[ \frac{\tau}{\tau^\prime}
\sinh \xi
+ \frac{g\cos\theta}{p_t \tau^\prime} 
\int_{\tau^\prime}^ {\tau} 
d\tau^{\prime\prime} E(\tau^{\prime\prime}) \right]
\end{equation}

It is thus clear that
$f$ does not depend on $\phi$, so the
$\partial \over \partial\phi$ term contributes nothing to the 
transport equation. However the $\theta$ dependence is 
still involved
through out the formalism which shows the non-Abelian effects.
The corresponding distribution $\bar{f}$ for antiquark can be 
written by just changing
$g$ to $-g$ in $f$ as can be checked out easily.

The transport equation has to be supplemented with the 
conservation constraint

\begin{equation}
\partial_\mu T^{\mu\nu}_{mat} + 
\partial_\mu T^{\mu\nu}_{YM}=0,
\label{thirteen}
\end{equation}

since it is the field energy that is being pumped in order to 
produce the $q \bar{q}$ pairs. 
More explicitly,

\begin{equation}
\partial_\mu T^{\mu\nu}_{mat} = -j_\mu^a F^{\mu\nu}_a \equiv
-\partial_\mu T^{\mu\nu}_{YM}
\label{thirteen}
\end{equation}

with

\begin{equation}
T^{\mu\nu}_{mat}=\int p^\mu p^\nu 
(f+\bar{f})d\Gamma d {\Omega}_Q
\label{fourteen}
\end{equation}

where the measures
$d{\Omega}_Q = \sin\theta d\theta d\phi$
, ${d{\Gamma} = {{\gamma d^3 {p}} \over {{({2 \pi})}^3 {p_0}}} 
= {{\gamma {p}_t d{p}_t d\xi} \over {({2 \pi})}^2}}$
and $ j^\mu_a =\int p^\mu Q_a (f-\bar{f} ) 
d\Gamma d {\Omega}_Q$. Here we have taken the value of the
degeneracy factor $\gamma=2$ corresponding to two flavours.
Since energy and momentum are conserved in each collision,
the moment
of the sum of the collision term vanishes:

\begin{equation}
\int p^\nu (C)d\Gamma d {\Omega}_Q=0
\label{sixteen}
\end{equation}

Now taking the first moments of the Boltzmann equation 
and integrating 
over the color degrees of freedom for $f$ and $\bar{f}$ 
and making use of
the conservation of energy and momentum, we obtain

\begin{equation}
\partial_\mu T^{\mu\nu}_f + gE(\tau ) 
\int d\Gamma d {\Omega}_Q p^\nu
\frac{\partial (f- \bar{f})}{\partial \xi}  
+ 2 \int d\Gamma d {\Omega}_Q S = 0 
\end{equation}

where 

\begin {equation}
T^{\mu\nu}_f = \mbox{diag} (E^2/2 ,E^2/2 ,E^2/2 ,-E^2/2 )
\end{equation}

is the energy momentum tensor for the field.
We may solve for the electric field by employing the 
same procedure as in
\cite {banerjee}, and by employing the symmetry 
$\bar{f}(\tau,\xi,p_T,\theta)=f(\tau,- \xi,p_T,\theta)$.
We thus obtain the equation governing the decay of the 
source field to be 

\begin{equation}
{ dE(\tau) \over d\tau }- \frac{2g\gamma }{2\pi} 
\int^\infty_0 dp_t p_t^2
\int^\infty_0 d\xi\sinh\xi \int^{\pi /2}_0 d\theta\sin\theta
[f - \bar{f} ] +\frac{4\pi}{7}\bar{a} \vert E(\tau) 
\vert^{3/2} = 0
\label{sixteen}
\end{equation}

where $\bar{a} = a\zeta (5/2)\exp (0.25/\alpha )$,  
$ a = c(g/2)^{5/2}\frac{\gamma }{(2\pi)^3}$ and  
$ c = \frac{1}{(4\pi)^3)}$.
Finally, $\zeta (5/2) =1.342$ is the Reimann zeta function.

Equations  (37) and (26) are as yet underdetermined since the 
local temperature $T(\tau)$ is free. 
In order to fix the form of $T(\tau)$, we appeal to the 
relaxation time
approach that we are employing, and assume 
that the the particle energy density differs 
negligibly from the 
equilibrium energy density.
We may then relate, by an ansatz, the proper energy 
density which is
defined by

\begin{equation}
\epsilon(\tau)=\int d\Gamma d {\Omega}_Q (p^\mu u_\mu)^2 
(f+\bar{f})
\end{equation}

to the temperature by its equilibrium value, whence,

\begin{equation}
T(\tau)=({15 \over (7 \gamma \pi^3)}{\epsilon(\tau)})^{1/4}.
\end{equation}

It may be mentioned that the weaker condition of
energy-momentum conservation that we have employed here
, in fact, satisfies the Yang-Mills equations
$ D_{\mu} \vec{F}^{\mu,\nu} = \vec{j}^{\nu} $ as well.

Finally, we pause briefly to discuss the effects of 
hard thermal loops
that have been emphasized recently \cite{h1}. 
They have been
derived from the classical transport equation as well by 
Kelly {\it et al}
\cite{kelly1,kelly2}. The latter derivation, which is 
important for us
here, is based on an analysis of the Vlasov equation which 
properly describes the expansion of an already 
equilibrated gas, in a
back-ground field but with 
with no source or sink. It is, therefore, necessary to 
determine whether
the more complete transport equation such as the one 
that we have here
will have any effect on the results of 
Ref. \cite{kelly1,kelly2}.
Conversely, it is also necessary to determine how the 
hard thermal loops
will affect the purely classical results that will be 
obtained here.
Let us recall that the derivation of 
Kelly et al \cite{kelly1,kelly2}
consists of a systematic expansion of the 
distribution function as well
as the background field term in powers of the 
coupling constant. The
zeroth order term is merely the Fermi-Dirac term for the 
quarks. Since we
are solving here the approach to the equlibrium, 
there is necessarily a
dependence of the distribution functions and other 
physical quantities on
$g$. It remains to disentangle the contribution 
coming from the hard
thermal loops. Indeed, observe that the two extra 
terms that we have at
hand here are of higher order in $g$. First of all,
the source term is
non-analytic and does not even admit an expansion 
in powers of $g$
\cite{fnote1}. The other collision term is easily seen 
to be of 
order $g^4$ or higher. We thus conclude that the 
conclusions of 
Ref. \cite{kelly1,kelly2} remain unaffected, and 
we may take over those
results {\it in toto}, supplementing our classical results.
	
\section{COMPUTATIONAL PROCEDURE}	
We have adopted here a double-self consistent method to 
determine $(f,\epsilon , ...)$ ,following
the work of Banerjee et al \cite{banerjee} in the Abelian case.
The procedure follows the scheme $\{T(\tau)_{trial},\, 
E(\tau)_{trial}\}    
\, \, \rightarrow \{f, \bar{f} \, \, E(\tau)\} \, \, 
\rightarrow
\{f,\bar{f}\} \, \, \rightarrow T(\tau) \, \, \rightarrow $ 
...
by repeated use of equations (20), (28), (20), (30). The
iteration terminates as soon as a convergence is established
in the solutions for $E(\tau), T(\tau)$. All the desired
quantities are thereby consistently determined.

\section{Results and discussions}
 Before we proceed to present  and discuss the results, 
  a few comments about the choice of the value of the
 parameters in the computation.  
 We put $g=4$ throughout our calculations.
 Since lattice computation results predict a phase
 transition from the baryonic phase to the quark-gluon
 phase at densities ${\sim 5-10 \, GeV}$, it has also
 been customary to take an initial energy density in the
 same range in the Flux tube model. However, greater
 care needs to be taken before the initial energy
 densities in URHIC are chosen. Indeed, a fairly reliable
 estimate of the time required for 
 the partons to be
 produced in central collisions after the two nuclei have
 suffered the maximum overlap is $\sim 0.05 \, - 0.1 \,
 fm$; a simple dimensional analysis leads to a value of
 the initial energy density to be $(1/2) E_0^2 \approx 
 500 \, -  1000 \, GeV/fm^3$, the precise value depending 
 on the the magnitude 
 of $g$ in 
$\sqrt{g E} = {1 \over \tau_0}$. 
Since in the colour plate model, we set the zero of time 
not at maximum overlap, but at the instant when the
plates have completely crossed each other, we shall take
the initial energy density $\epsilon= 300 GeV/fm^3$.
We shall also be guided by the results of Geiger and
Kapusta \cite{geiger} in our choice of the values of
$\tau_c$, and take $\tau_c = 0.2 fm$, to be a realistic
value \cite{fnote2}. The scale for the
hydrodynamic limit will be set by the choice for $\tau_c$
as well as the formation time $\tau_0$. A typical value
for the analysis at hand is 0.001 fm, which we employ
here. Finally, we study the other extreme case, the
collisionless limit, by the choice $\tau_c=5 fm$. It
is instructive to compare how the realistic regime
behaves in relation to the results that we obtain for
the two extreme limits.

We have studied the decay of the chromoelectric field,
the evolution of the particle energy and number
densities, the evolution of the distribution function
and its approach to the equlibrium state. We have compared
them with the corresponding Abelian results. In
addition, we have also evaluated a quintessentially
non-Abelian quantity, {\it viz}, the expectation value of the
angle that the quark charge makes with the 
direction of the chromoelectric field.  In the
current model the chromoelectric field not only decays
to produce the $q\bar{q}$ pairs, but is itself built up
by the receding nuclei which act as colour plates. We
have, therefore, also studied the particle energy per
unit transverse area as a
function of ordinary time to highlight this feature.

\subsection{Discussion of the results}
 We shall present the results of our analysis in the 
 three regimes corresponding to the hydrodynamic, the
 realistic, and the collisionless cases. Comparison
 between the Abelian and the non-Abelian systems will be
 taken up subsequently.

Consider the hydrodynamic limit first. The physical
quantity of utmost importance is
the chromoelectric field, whose dynamics
is most readily determined in the Flux tube model at
hand. Indeed, apart from determining
the production and the acceleration of 
the partons in the Flux tube model at hand, it has an 
additional role which has been emphasized by Svetitsky
\cite{svetitsky}: the charms which
are
produced in the pre-equilibrium stage would not only
interact with the gas of light quarks and gluons,
before either forming a $ J/{\psi}$ or an open charm
mesonic state, but will also be
influenced by the background field. 
Thus the study the evolution of the mean electric
field in the pre-equilibrium stage acquires an added 
importance. It is a merit of the Flux tube model that
we can readily determine the evolution of the
background electric field. Note that in contrast, no
such information can be extracted in the more
microscopic models such as that of Geiger and Kapusta
\cite{geiger}.
 In fig. 1 the decay of the field is shown. 
 Recall that $\tau_c = .001 fm$.
As can be seen from Fig. 1, the electric field has hardly
decayed at all, 
with a percentage decay less than 2 \%, even at
$\tau=1.5 fm$. The corresponding particle energy
density, shown in Fig. 4 is also negligibly small, with
the ratio with respect to the initial energy
density being $\sim 10^{-5}$. The Abelian situation
is an order of magnitude better, but is still
hopelessly small. And indeed, the number density is
also very small, and as 
shown in Fig. 7 gets saturated at $\sim .05 / fm^3$.
and the temperature also stabilizes to a value $\sim 40
MeV$, which is much less than the temperatures
required. Clearly, it is very unlikely that the plasma
does not go through a prequilibrium phase.

 The behaviour of the system for $\tau_c = 0.2$ and $5
 fm$ is in sharp contrast to the case discussed above.
 There is a significant decay of the electric field,
 which is shown in Figs. 2 and 3 for the two
 respective values of $\tau_c$.  The electric field
 decays  by about 15\% of the original
 value, at $\tau=1.5 fm$. The corresponding energy
 densities are also not very different, and yield
 $\sim 10-15\%$ of the initial field energy density
 (see Figs. 5 and 6). The corresponding temperatures
 are $\sim 300 MeV$, which is quite realistic. Note
 that the parton cascade model \cite{geiger} predicts
 a value about 300 MeV at $2.5 fm$.  The real difference
 between the collisionless case and the "realistic"
 case is in the number density. Indeed as Figs. 8 and
 9 show, the plasma produced for $\tau_c=0.2 fm$ has a
 number density $\sim 15-20/fm^3$, the corresponding
 number for $\tau_c=5 fm$ is three to four times
 smaller. In other words, the system is more dense in
 the former case than the latter. Clearly, this
 distinction should show up in signatures such as
 dilepton production which are sensitive to both the
 number as well as the energy density. 

 It thus follows that
 the flux tube model, with the incorpration of the
 colour degrees of freedom, definitely rules out
 instantaneous equilibration as envisaged in the
 approach of Baym \cite{baym}; further, it also
 distinguishes the collisionless limit from realistic
 values of equilibration time.

  Finally, it should be emphasized that the flux tube
  model as employed here not only pumps in the field
  energy to particle production, but also contributes
  to the field energy by virtue of the recession of
  the colour charged nuclei from each other. In fact,
  a crude estimate of the time required to convert all
  the plate energy to the field energy turns out to be
  $\sim 5 fm$  for $200 GeV/$nucleon. In any case, it is
  therefore misleading to interpret the ratios we have
  shown in Figs. 4-6 as the fraction of the total
  field energy that has gone into the particles. To
  emphasize this, we have evaluated the dependence
  of the field and  particle energies per unit area
  by integrating over the contribution
  along the longitudinal direction.
  The results are shown in Fig. 10,
  from which it is clear that there is a lot more
  energy in the field than in the particles. For the
  same reason, it is also misleading to interpret the
  field energy density at $\tau=0$ as the counterpart
  of the energy densities employed in lattice
  computations to study the transition from the
  hadronc to the QGP phase.

\subsection{Comparison with Abelian results}

It is clear from Figs. 1-9 that the Abelian and the
non-Abelian results bear little resemblence, belying
the expectation that the "Abelian dominance" holds in
this case. The difference is most dramatically
highlighted at $\tau_c= 0.2 fm$. Whereas almost all
the initial field has decayed in the Abelian case,
only 30\% has done so in the non-Abelian case.
Accordingly, The particle energy density is larger by
a factor of $\sim 5$ , the number density by a factor
of $\sim 4$, and yields an abnormally large
value of$\sim 800  MeV$ for the temperature. 
Earlier Abelian
calculations \cite{kajantie,banerjee,asakawa} yielded
reasonable values for the temperature because of
unrealistic values for the initial field energy. It is
a universal feature that the non-Abelian plasma is 
rarer and  cooler than its Abelian counterpart. There
are other significant features. Consider the
hydrodynamic case. It is seen that although the decay
in the non-Abelian field is more than the Abelian
field, the corresponding energy density is smaller than
in Abelian case. Indeed, in the hydrodynamic limit,
the Abelian ananlysis yields a temperature $\sim 200
MeV$, in contrast to $\sim 40 MeV$ in the non-Abelian
case. Thus the incorporation of the colour degree of
freedom rules out instantaneous hydrodynamic
evolution. Of course, a colourless plasma does not
have reasonable temperature for any other $\tau_c$.

Yet another interesting aspect that emerges from our
studies is in the close interplay between the value
of $\tau_c$ and the colour degree of freedom. 
 While the field decays faster in the non-Abelian case
  in the hydrodynamic limit, the trend reverses at
  $\tau_c =0.2 fm$ and gets restored in the
  collisionless limit. In contrast, the particle
  energy density is smaller for a coloured plasma in
  hydrodynamic limit. Further, it continues to be so
  at $\tau_c=0.2 fm$ but becomes larger than the
  Abelian case in the collisionless limit. These
  non-trivial manifestations of the colour charge may
  be expected to have an important bearing on the
  other bulk properties of the plasma.

 In order to gain some insight into the above
 features, we have also evaluated the expectation
 value of $\Theta \equiv <cos^2\theta>$ which 
 yields the rms
 value that the particle charge makes with the field
 direction in the colour space. Note that the above
 quantity is gauge invariant, and hence physical. Fig.
 11 shows $\Theta(\tau)$ for $\tau_c=0.2 fm$. It may
 be seen that the value saturates around 0.25,
 corresponding to $\theta \sim \pi/3$. 
 The corresponding value in the non-Abelian case is
 strictly zero.
 The effect of the
 background field is thus reduced, leading to
 dominance of the collision term in the expansion of
 the particles. Indeed, it shows up most clearly in
 the approach to the equilibrium, where one now
 expects that the plasma equilibrates the fastest 
 in the direction
 normal to that of the field (in the colour space).
 This is corroborated as may be seen in Fig. 12.

\section{CONCLUSION}

 To conclude, we have studied the production and the
 equilibration of a genuinely non-Abelian plasma
 with the colour degree of freedom incorporated in both
 the source and the background field term in the
 transport equation. In the $SU(2)$ gauge theory that
 we have considered here, the distribution function is
 defined in the extended phase space.
 We find that
 this approach recaptures in an elegant manner many of
 the findings of the more microscopic parton cascade
 model. It has the further advantage that it indeed
 exhibits the colour degree of freedom manifestly, and
 allows us to compute various gauge invariant
 quantities. Significantly, we also find that the
 Abelian approximation, employed hitherto in most
 studies of equilibration of QGP is rather too drastic
 to be used for any quantitative analysis and
 comparison with the experiments. The study also almost
 rules out instantaneous equilibration, and also
 strongly suggests that the collisionless limit may
 also not be the favoured in URHIC.

 To be sure, we have not made any comparison with the
 experimental findings here, for the simple reason that
 we are as yet dealing with a simpler gauge group , and
 ignored the gluonic component altogether. The
 indications from the present study are unmistakable,
 though. Indeed, the particle production is enhanced
 because of increase in the phase space available, and
 for the same reason the plasma will be cooler than the
 Abelian counterpart. The energy goes to the colour
 degree of freedom, and does not simply heat the system
 as it would happen in a colourless plasma. If we
 consider the realistic $SU(3)$ case, this feature will
 get further accentuated; For the same initial
 configuration, we may expect a rarer and a cooler
 plasma. Of course, there are other features which are
 intrinsic to $SU(3)$: there is yet another Casimir
 invariant, and the gluon term will also have to be
 incorporated. These are under study at the present and
 will be reported else where, with a full discussion of
 the signatures of the QGP in the flux tube model.

{\bf Acknowledgment:}
It is our great pleasure to thank Rajeev Bhalerao who
 made available the program for the Abelian studies
 which was extended for the present case. We also
 recall with pleasure the very many discussions that we
 have had with him.

        \newpage

\subsection*{Figure captions}
\noindent
{\bf FIG.~1.} Decay of the chromoelectric field as a function
of proper time (in units of fermi), for $\tau_c=.001 fm$. 
The solid line refers to the non-Abelian
	      case, and the broken line to the Abelian case.

\noindent
{\bf FIG.~2.} Decay of the chromoelectric field as a
function of proper time (in units of fermi), for 
$\tau_c=.2 fm$. The solid line refers to the non-Abelian
case, and the broken line to the Abelian case.

\noindent
{\bf FIG.~3.} Decay of the chromoelectric field as a
function of proper time (in units of fermi), for 
$\tau_c=5 fm$. The solid line refers to the non-Abelian
case, and the broken line to the Abelian case.

\noindent
{\bf FIG.~4.} The particle energy density scaled w.r.t the
initial field energy density as a
function of proper time (in units of fermi), for 
$\tau_c=.001 fm$. The solid line refers to the non-Abelian
case, and the broken line to the Abelian case.

\noindent
{\bf FIG.~5.} The particle energy density scaled w.r.t the
initial field energy density as a
function of proper time (in units of fermi), for 
$\tau_c=.2 fm$. The solid line refers to the non-Abelian
case, and the broken line to the Abelian case.

\noindent
{\bf FIG.~6.} The particle energy density scaled w.r.t the
initial field energy density as a
function of proper time (in units of fermi), for 
$\tau_c=5 fm$. The solid line refers to the non-Abelian
case, and the broken line to the Abelian case.

\noindent
{\bf FIG.~7.} The particle number density as a
function of proper time (in units of fermi), for 
$\tau_c=.001 fm$. The solid line refers to the non-Abelian
case, and the broken line to the Abelian case.

\noindent
{\bf FIG.~8.} The particle number density as a
function of proper time (in units of fermi), for 
$\tau_c=.2 fm$. The solid line refers to the non-Abelian
case, and the broken line to the Abelian case.

\noindent
{\bf FIG.~9.} The particle number density as a
function of proper time (in units of fermi), for 
$\tau_c=5 fm$. The solid line refers to the non-Abelian
case, and the broken line to the Abelian case.

\noindent
{\bf FIG.~10.} The particle energy/unit transverse area 
(solid line)
and the field energy/unit transverse area (dashed line)
as a function of time (in fermi). The energy densities are in
$GeV/fm^2$ and $\tau_c=.2 fm$.

\noindent
{\bf FIG.~11.} $<cos^2\theta>$ as a function of proper time
(in fermi) at $\tau_c=.2 fm$.

\noindent
{\bf FIG.~12.}  $f/f_{eq}$ as a function of proper time at 
   $p_t=200 MeV$, $\xi=0$ and $\tau_c=0.2 fm$ for three
   different angles corresponding to $cos\theta=0$ 
   (solid line), $cos\theta =.25$ (dash line just below
   the solid line), and $cos\theta=1$ (the other dash line).
   Note that the equlibration is fastest at $\theta=\pi/2$.


\begin{thebibliography}{99}
\bibitem{fnote1} A perturbative source term which would
       arise for time dependent fields will not alter the
       situation since it is of order $g^2$.
\end{thebibliography}
\end{document}